\documentclass[epjST_arXiv]{svjour_arXiv}
\usepackage{graphics}
\usepackage{amsmath}
\usepackage{color}
\usepackage{dsfont}

\newcommand{\average}[1]{\left \langle #1 \right \rangle}

\begin{document}
\title{On the robustness of strongly correlated multi-photon states in frustrated driven-dissipative cavity lattices}
\author{Riccardo Rota\inst{1}\fnmsep\thanks{\email{riccardo.rota@univ-paris-diderot.fr}}
\and Wim Casteels \inst{2}\fnmsep\thanks{The first two authors contributed equally to this work.}  
\and Cristiano Ciuti\inst{1}}
\institute{Universit\'e Paris Diderot, Sorbonne Paris Cit\'e, Laboratoire Mat\'eriaux
	et Ph\'enom\`enes Quantiques, CNRS-UMR7162, 75013 Paris, France \and
	TQC, Universiteit Antwerpen, Universiteitsplein 1, B-2610 Antwerpen, Belgium}

\abstract{
We present a theoretical study on the robustness of multi-photon states in a frustrated lattice of coupled nonlinear optical cavities, which are described by a driven-dissipative Bose-Hubbard model. In particular, we focus here on a Lieb lattice with two elementary cells and periodic boundary conditions. Due to the geometric frustration of the lattice, the non-equilibrium steady state can exhibit dark sites with low photon density and strong correlations, ascribable to the population of multi-photon modes. We explore the sensitivity of such strong correlations on the random inhomogeneity of the lattice parameters. We show that the correlations are more sensitive to the inhomogeneity of the cavity frequencies than to the random fluctuations of the hopping strength.} 
\maketitle
\section{Introduction}\label{sec:intro}


Since several years, it is possible to study quantum many-body systems on lattice geometries using ultracold gases, confined in periodic potential formed by standing waves of laser light \cite{BlochReview}. Despite the diluteness of the particles in these systems, strongly correlated states may arise due to the interplay between the delocalization of the atoms over the lattice and their mutual interaction \cite{Greiner02}.


In presence of a geometric frustration of the lattice, the energy spectrum may display non-dispersive flat bands, corresponding to states localized within a finite volume of the lattice. 
Flat energy bands occur in a large variety of systems, from graphene \cite{Nakada96} to quantum magnets \cite{Ramirez94} and unconventional supercondutors \cite{Schnyder11}, and they lead to the emergence of nontrivial correlated structures, for example in the context of spin ice \cite{Morris09,Nisoli13}, quantum Hall systems \cite{Neupert11,Bergholtz13,Parameswaran13,DePicciotto97,Tang11,Petrescu} and Josephson junctions \cite{Sigrist95,Feigelman04}.

In the context of ultracold atoms, the problem of condensation in flat bands has been extensively studied from the theoretical point of view \cite{Wu07,Huber10,Apaja10,Tovmasyan13}, but its experimental investigation still presents important drawbacks. Although the high degree of control in the manipulation of the confining optical potential has allowed to load quantum gases in frustrated lattices \cite{Oschlager11,Wirth11,Soltan-Panahi12a,Struck11,Soltan-Panahi12b,Tarruell12,Jo12,Windpassinger13}, it is not easy to access the flat bands since these typically appear at energies far above the ground-state \cite{Aidelsburger15,Taie15}.

An alternative approach is to consider photons propagating in arrays of coupled nonlinear cavities, experimentally realizable with semiconductor micropillars \cite{Weisbuch,DeveaudBook} or superconducting circuits at microwave frequencies \cite{Schoelkopf08,You11}. In regimes of strong coupling between light and matter, the photons hybridize with the exciton modes of the nonlinear media, giving rise to the so-called polaritons. The mixed light-matter nature of these quasi-particles allows for relevant scattering channels for the photons and yields an effective interaction among them, similar to that of ultracold atoms in quantum gases \cite{CarusottoRMP}.

An important aspect of the photonic systems, which represent a crucial difference with respect to atomic gases, is related to their nonequilibrium nature. Due to the finite transmittance of the mirrors of optical cavities, these systems are unavoidably subjected to photon losses. Therefore, a continuous drive has to be applied to the cavity in order to inject photons and reach a non-trivial steady state. This feature, nevertheless, allows for the advantageous possibility to engineer the coupling with the environment and to drive the system to a steady state of interest \cite{Diehl08}.

\begin{figure}
\begin{center}
\resizebox{0.58\columnwidth}{!}{%
\includegraphics{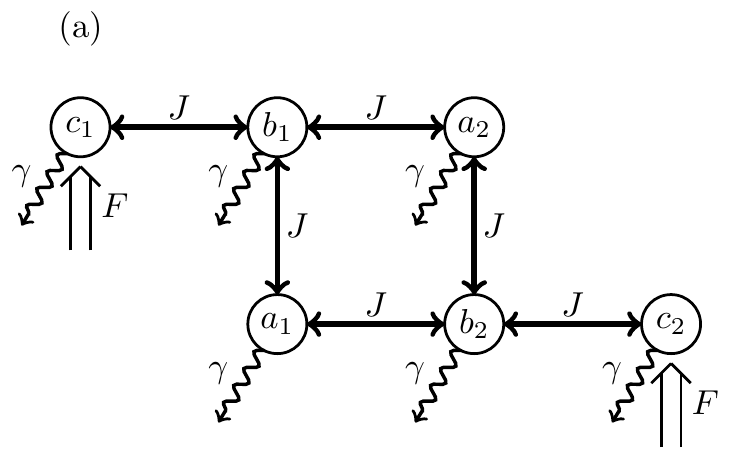} }
\resizebox{0.35\columnwidth}{!}{%
\includegraphics{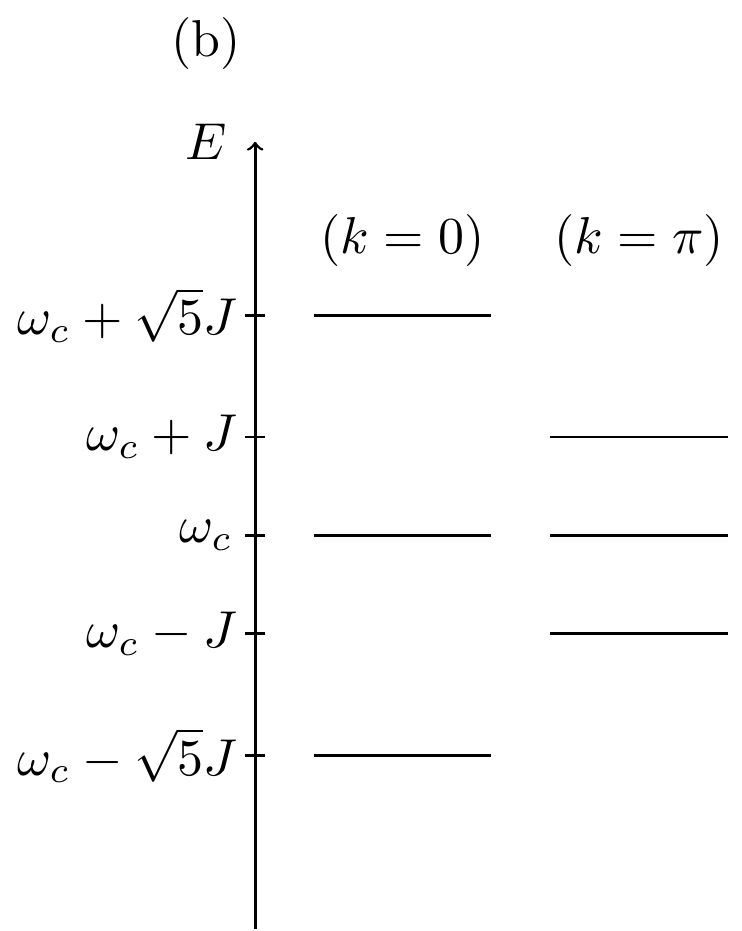}}
\end{center}
\caption{(a) Sketch of the driven-dissipative lattice studied in this work. The coherent driving is applied only to the $c$-sites. (b) Single-particle energy level structure for the closed non-interacting system. The levels are organized in two columns, according to the value of the wave-vector $k$ of their relative eigenstate.}
\label{fig:sketch}       
\end{figure}

In this paper, we study a simple driven-dissipative photonic system made up of six coupled optical cavities, arranged with the same geometry of a 1D Lieb lattice with two unit cells and periodic boundary conditions (see Figure \ref{fig:sketch}-(a)). Experimental realizations of extended Lieb lattices have been realized both with photonic systems \cite{GuzmanSilva14,Vicencio15,Mukherjee15,Baboux16} and ultracold gases \cite{Taie15}, showing the emergence of the localized states associated to the flat bands. Furthermore, theoretical studies of driven-dissipative Lieb lattices have shown that incompressible states with short-range crystalline order can arise in photonic systems with large nonlinearity (i.e. a strong effective photon-photon interaction)\cite{Biondi15}. 

In a recent work \cite{Casteels16Lieb}, it has been shown that, even in regimes of small nonlinearities, it is possible to create strongly correlated photonic modes and to probe them at the  dark sites of the frustrated lattice, where the mean photon population is largely reduced due to interference effects. Here, we want to study the behavior of these correlated states when the lattice presents a certain amount of disorder, focusing on a simple 1D Lieb lattice made of two unit cells. The moderate size of this system allow us to tackle the problem using a rather simple algorithm, based on an efficient truncation scheme of the correlation functions at high order \cite{Casteels16Truncation}. The results obtained show that correlations tends to disappear when the inhomogeneity on the cavity frequencies increases, while the correlated states are particularly robust in presence of disorder in the hopping between the cavities.

The paper is organized as follows: in sec. \ref{sec:methods} we discuss the theoretical framework describing the quantum system and the numerical method used; in sec. \ref{sec:results} we presents our results and in sec. \ref{sec:conclusions} we draw our conclusions.

\section{Theoretical model}\label{sec:methods}
The properties of interacting photons in cavity lattices can be accurately described in terms of a driven-dissipative Bose-Hubbard model. For the particular lattice in Fig. \ref{fig:sketch}-(a), one can consider the Hamiltonian ($\hbar = 1$ in the following):
\begin{eqnarray}\label{eq:Hamiltonian}
\hat{H} & = & \sum_{\substack{i=1,2 \\ s \in \{a,b,c\}}}\left(\omega_c\hat{s}_i^\dagger\hat{s}_i+\frac{U}{2}\hat{s}_i^\dagger\hat{s}_i^\dagger\hat{s}_i\hat{s}_i \right)  -J\sum_{i=1}^2 \left(\hat{a}_i^\dagger\hat{b}_i+\hat{b}_i^\dagger\hat{c}_i+ \hat{a}_i^\dagger\hat{b}_{3-i} + h.c.\right) \nonumber \\ 
& &+ \sum_{i=1}^2\left( Fe^{-i\omega_pt}\hat{c}_i^\dagger+F^*e^{i\omega_pt}\hat{c}_i \right) \ . 
\label{Ham} 
\end{eqnarray}
The operators $\{\hat{a}_i,\hat{b}_i,\hat{c}_i\}_{i=1,2}$ are the annihilation operators for photons on the different sites. The first term describes the on-site part with the resonator frequency $\omega_c$ and a photon-photon interaction strength $U$. The second terms represents the hopping between different sites, $J$ being the corresponding coupling. Finally, the last line represents a coherent drive on the $c$-sites, with amplitude $F$ and frequency $\omega_p$.

For the case of the closed non-interacting system ($F=0$, $U = 0$), the Hamiltonian in Eq. \ref{eq:Hamiltonian} can be diagonalized in reciprocal space, through the definition of the annihilation operators of the $k$-modes: $\hat{s}_k = 1/\sqrt{2} \sum_{j=1}^2 e^{i k j} \hat{s}_j$, where $k$ can assume only the two values $k=0$ and $k = \pi$. The single-particle energy spectrum contains six eigenstates (three for $k = 0$ and three for $k = \pi$, see Fig. \ref{fig:sketch}-(b)). A flat band at energy $E = \omega_c$ can be seen and the two eigenstates corresponding to it are
\begin{eqnarray}
|\Psi_{FB}^{(k = 0)}\rangle & =  & \frac{1}{\sqrt{10}} \left( | 1,0,0,0,0,0 \rangle - 2 | 0,0,1,0,0,0 \rangle
+ | 0,0,0,1,0,0 \rangle  -2 | 0,0,0,0,0,1 \rangle \right) \nonumber \\
|\Psi_{FB}^{(k = \pi)}\rangle & =  & \frac{1}{\sqrt{2}} \left( | 1,0,0,0,0,0 \rangle -  | 0,0,0,1,0,0 \rangle \right) \ ,
\end{eqnarray}
where we used the Fock basis representation $|n_{a_1},n_{b_1},n_{c_1},n_{a_2},n_{b_2},n_{c_2} \rangle$, with $n_s$ the number of photons on the $s$-site. It is easy to notice that the states corresponding to the flat band have zero occupancy on the sites $b_1$ and $b_2$, which are denoted as dark sites. 

The driven-dissipative dynamics is described by the Lindblad master equation for the time evolution of the density matrix $\hat{\rho}(t)$, given by:
\begin{equation}
\frac{\partial\hat{\rho}}{\partial t}= i\left[\hat{\rho},\hat{H}\right] + \frac{\gamma}{2} \sum_{i,s}\left(2\hat{s}_i\hat{\rho}\hat{s}^\dagger_i - \hat{s}_i^\dagger\hat{s}_i\hat{\rho}-\hat{\rho}\hat{s}_i^\dagger\hat{s}_i \right),
\label{eq:Master}
\end{equation}
where $\gamma$ is the cavity loss-rate. The steady-state properties of the system are determined by solving the master equation for $\frac{\partial\hat{\rho}}{\partial t}=0$. For sake of simplicity, we consider a reference frame rotating with the frequency of the pump $\omega_p$. By performing this change of the reference frame, the Hamiltonian becomes time-independent and the relevant parameter is the detuning between the pump and the cavity frequencies, namely $\Delta = \omega_p - \omega_c$.

In order to perform the numerical integration of the master equation Eq. (\ref{eq:Master}), the full Hilbert space, with an infinite dimension, has to be truncated. Many standard truncation schemes rely on a cutoff on the number of photons per site which results in a complexity that grows exponentially with the number of sites. They quickly become inefficient for lattices of moderate size, like the one we are considering in the present work.

Instead we use a numerical approach based on a global truncation scheme at the level of the correlation functions \cite{Casteels16Truncation}. 
From the master Eq. (\ref{eq:Master}) we find the equations of motion for the correlation functions $C = \langle\prod_{i,s}\hat{s}_i^{\dagger n_{i,s}}\hat{s}_i^{m_{i,s}}\rangle$, with $\{n_{i,s}\}$ and $\{m_{i,s}\}$ non-negative integer numbers:
\begin{eqnarray}
\frac{\partial C}{\partial t} & = &\sum_{i,s}\left(-i\Delta\left(n_{i,s} - m_{i,s}\right)+\frac{\gamma}{2}\left(n_{i,s} + m_{i,s}\right)  \right)C \nonumber\\
&+& i \frac{U}{2}\sum_{i,s}\left(\left[n_{i,s}\left(n_{i,s} - 1\right) - m_{i,s}\left(m_{i,s} - 1\right)\right]  \right)C + iU\sum_{i,s}\left(n_{i,s} - m_{i,s}\right) C\lvert_{n_{i,s}+ \atop m_{i,s}+} \nonumber \\
&-& i J\sum_{<(i,s),(j,r)>}\left(n_{i,s}C\lvert_{n_{j,r}+ \atop n_{i,s} -}  - m_{i,s}C\lvert_{m_{j,r}+ \atop m_{i,s}-} + n_{j,r}C\lvert_{n_{i,s}+ \atop n_{j,r}-}  - m_{j,r}C\lvert_{m_{i,s}+ \atop m_{j,r}-} \right) \nonumber \\
&+& i\sum_{i} \left(n_{i,c}F^*C\lvert_{n_{i,c}-}-m_{i,c}FC\lvert_{m_{i,c}-}\right),
\label{eq:CorrFun}
\end{eqnarray}
where we used the following notation: 
\begin{eqnarray}
\label{eq:defineCpm}
C\lvert_{n_{i,s}\pm} &  =  & \langle \hat{s}_i^{\dagger (n_{i,s} \pm 1)}\hat{s}_i^{m_{i,s}}\prod_{(k,t) \neq (i,s)}\hat{t}_k^{\dagger n_{k,t}}\hat{t}_k^{m_{k,t}} \rangle \ , \nonumber \\
C\lvert_{n_{j,r}\pm \atop n_{i,s} \pm} &  =  & \langle \hat{r}_j^{\dagger (n_{j,r} \pm 1)}\hat{r}_j^{m_{j,r}} \hat{s}_i^{\dagger (n_{i,s} \pm 1)}\hat{s}_i^{m_{i,s}} \prod_{(k,t) \neq (j,r) \atop (k,t) \neq (i,s)}\hat{t}_k^{\dagger n_{k,t}}\hat{t}_k^{m_{k,t}} \rangle \ ,
\end{eqnarray}
with the indices $i,j,k \in \{Ê1,2\}$ denoting the unit cell of the lattice and the indices $r,s,t \in \{Êa,b,c\}$ denoting the site. In the definition of the correlation function $C\lvert_{n_{i,s}\pm}$, the product runs over all the sites of the lattice except the site $(i,s)$; similarly, in the definition of the correlation function $C\lvert_{n_{j,r}\pm \atop n_{i,s} \pm} $, the product runs over all the sites different from $(i,s)$ and $(j,r)$.

The notation $<(i,s),(j,r)>$ in (\ref{eq:CorrFun}) denotes that the sum runs over all the sites $(i,s)$ and $(j,r)$ that are coupled through the hopping terms in the Hamiltonian (\ref{Ham}). Since the Hamiltonian (\ref{Ham}) is nonlinear this leads to an infinite hierarchy of coupled equations for the correlation functions, in which the solution for a given $C$ depends on the solution for other correlation functions at different order. In order to proceed we have to introduce a truncation scheme, which reduces the infinite hierarchy to a finite set of equations that can be solved numerically. The simplest possibility corresponds to setting all correlation functions equal to zero of which the correlation order $\max(\sum m_{i,s}, \sum n_{i,s})$ is larger than a cutoff value $N_c$. This is also known as a global 'hard' cutoff and corresponds to an expansion around the vacuum \cite{Casteels16Truncation}. Convergence of the results is then verified by increasing the value of the cutoff $N_c$. 
Considering the moderate size of the lattice we want to study, this approach is ideally suited because of its easy implementation and the low computational effort demanded in the calculations.

The reliability of this approach has been also verified by comparing the results for a chosen set of parameters with those obtained with the corner-space renormalization method \cite{Finazzi15}. 
This method provides a wise procedure to select the relevant subspace (i.e. the corner) of the Hilbert space of an extended driven-dissipative lattice, where the full Master Equation for the density matrix (Eq. \ref{eq:Master}) can be solved. The states spanning the corner are determined through an iterative procedure, using eigenvectors of the steady-state density matrix of smaller lattice systems, merging in real space two lattices at each iteration and selecting a bipartite basis of product states by maximizing their joint probability. Increasing the dimension of the corner space, one recovers more and more accurate results for the steady-state density matrix, until convergence within the desired accuracy is reached. This possibility allows thus to keep easily under control the errors due to the particular choice of the corner and makes the corner-space renormalization method a good reference to test the results for the correlation functions obtained within the truncation method proposed in Ref. \cite{Casteels16Truncation}.

\section{Results}\label{sec:results}

\begin{figure}
\begin{center}
\resizebox{0.7\columnwidth}{!}{
\includegraphics{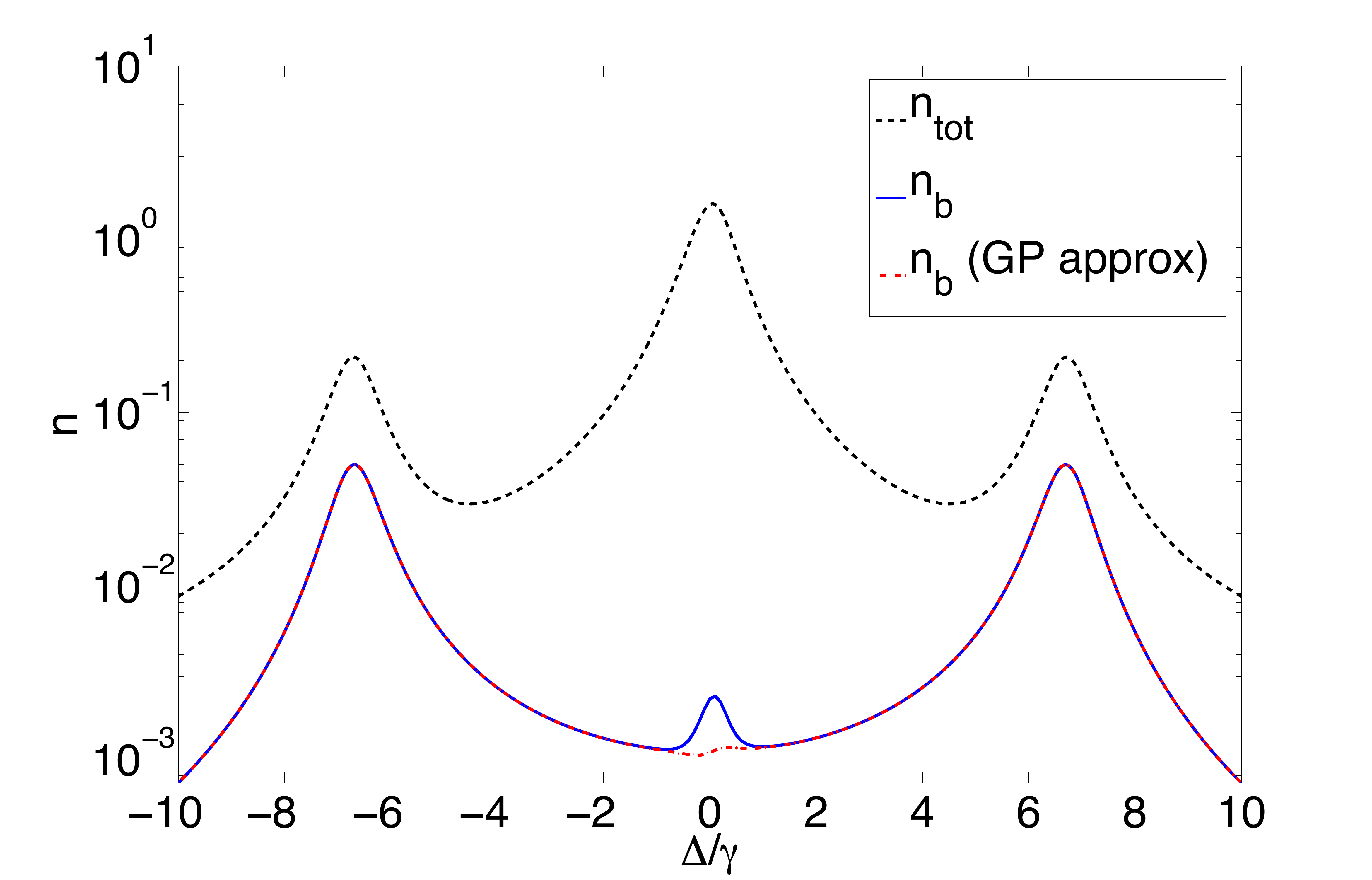} 
}
\end{center}
\caption{Total photon density $n_{tot}$ (black dashed line) and mean photon density $n_b$ on the $b$-sites (solid blue line) as a function of the detuning $\Delta$ (in units of $\gamma$), for $U = 0.1 \gamma$, $J = 3 \gamma$ and $F = 0.5 \gamma$. The results for $n_b$ obtained with the Gross-Pitaevskii approximation is also shown (red dash-dotted line).}
\label{fig:populationdark}       
\end{figure}

We start our discussion considering the photon density on the lattice obtained at the steady state. In Fig. \ref{fig:populationdark} we plot the total density on the whole lattice $n_{tot} = \average{\sum_{i,s} \hat{s}_i^\dagger \hat{s}_i}$ and the mean density at the $b$-sites $n_b = \average{\hat{b}_1^\dagger \hat{b}_1} = \average{\hat{b}_2^\dagger \hat{b}_2}$ as a function of the detuning $\Delta$. For the total density $n_{tot}$ we can clearly see three peaks, for $\Delta = \pm \sqrt{5} J$ and $\Delta = 0$. At these values, the driving is at resonance with the single-particle modes at $k = 0$ (see fig. \ref{fig:sketch}-(b)) and injects a large number of photon in the lattice. The density $n_b$ shows instead a different behavior: it presents two large peaks for $\Delta = \pm \sqrt{5} J$, as the driving is at resonance with the states of the dispersive bands. For $\Delta = 0$, the density $n_b$ is orders of magnitude smaller than the total density on the lattice, indicating that in this regime the $b$-sites are dark with respect to the total density. 
Furthermore, around $\Delta = 0$, there is a small peak in the dark site density which is attributed to the presence of correlated photons. Indeed, this feature is not present for the linear system (with $U=0$) and it is not captured by the semiclassical Gross-Pitaevskii prediction (red dash-dotted line in Fig. \ref{fig:populationdark}) which neglects all correlations.

\begin{figure}
\begin{center}
\resizebox{0.42\columnwidth}{!}{%
 \includegraphics{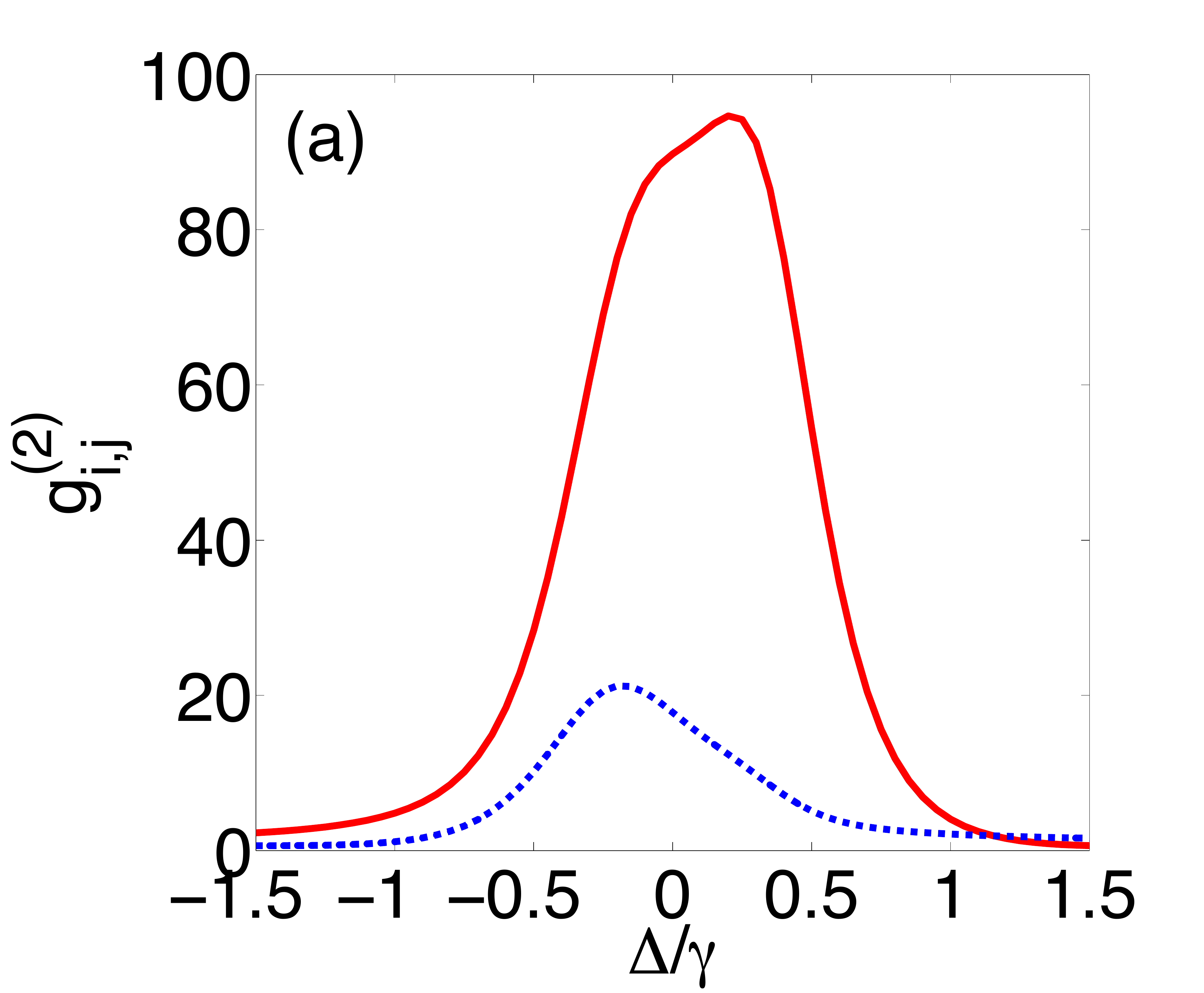} }
 \resizebox{0.42\columnwidth}{!}{%
 \includegraphics{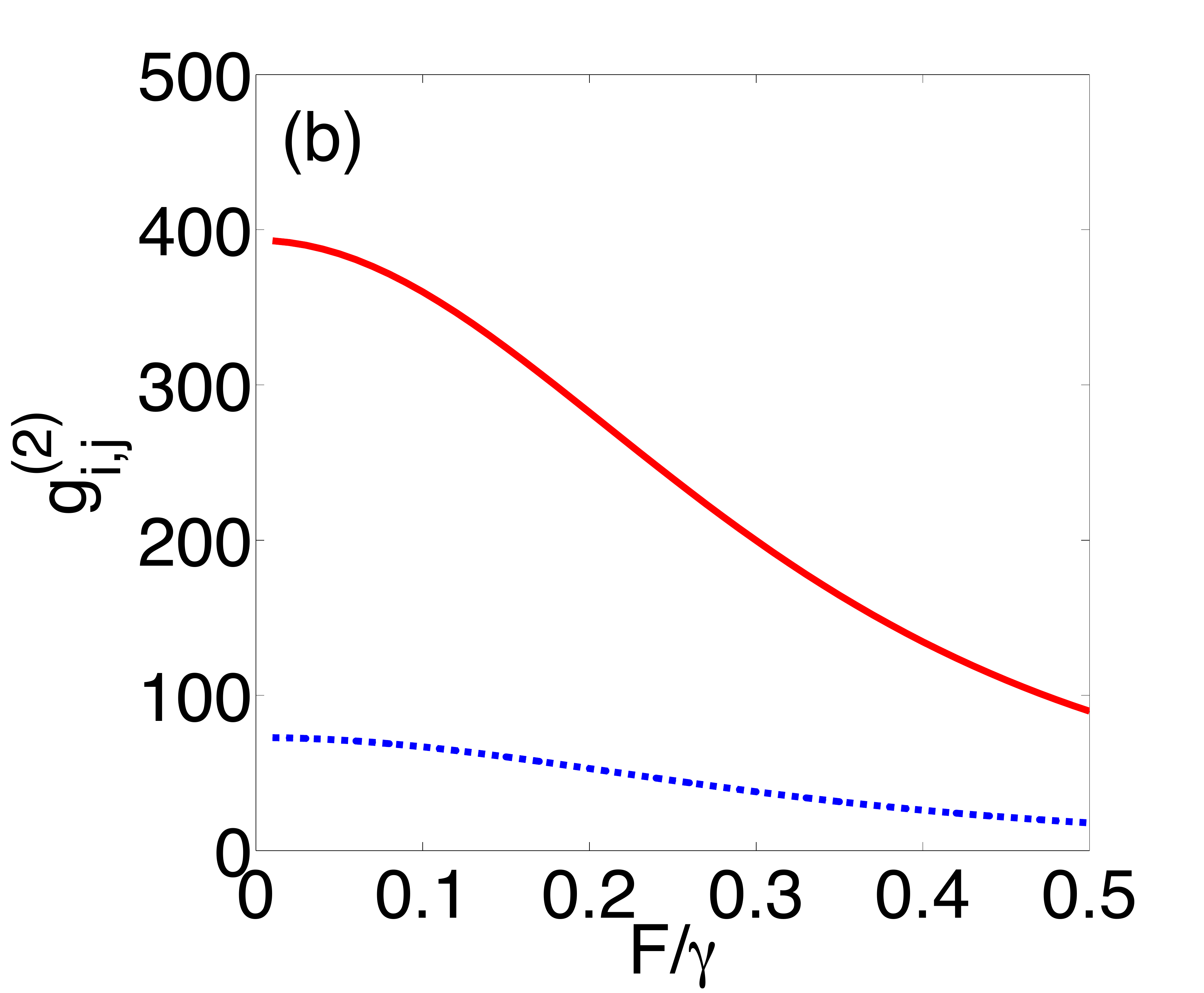} }\\
 \resizebox{0.42\columnwidth}{!}{%
 \includegraphics{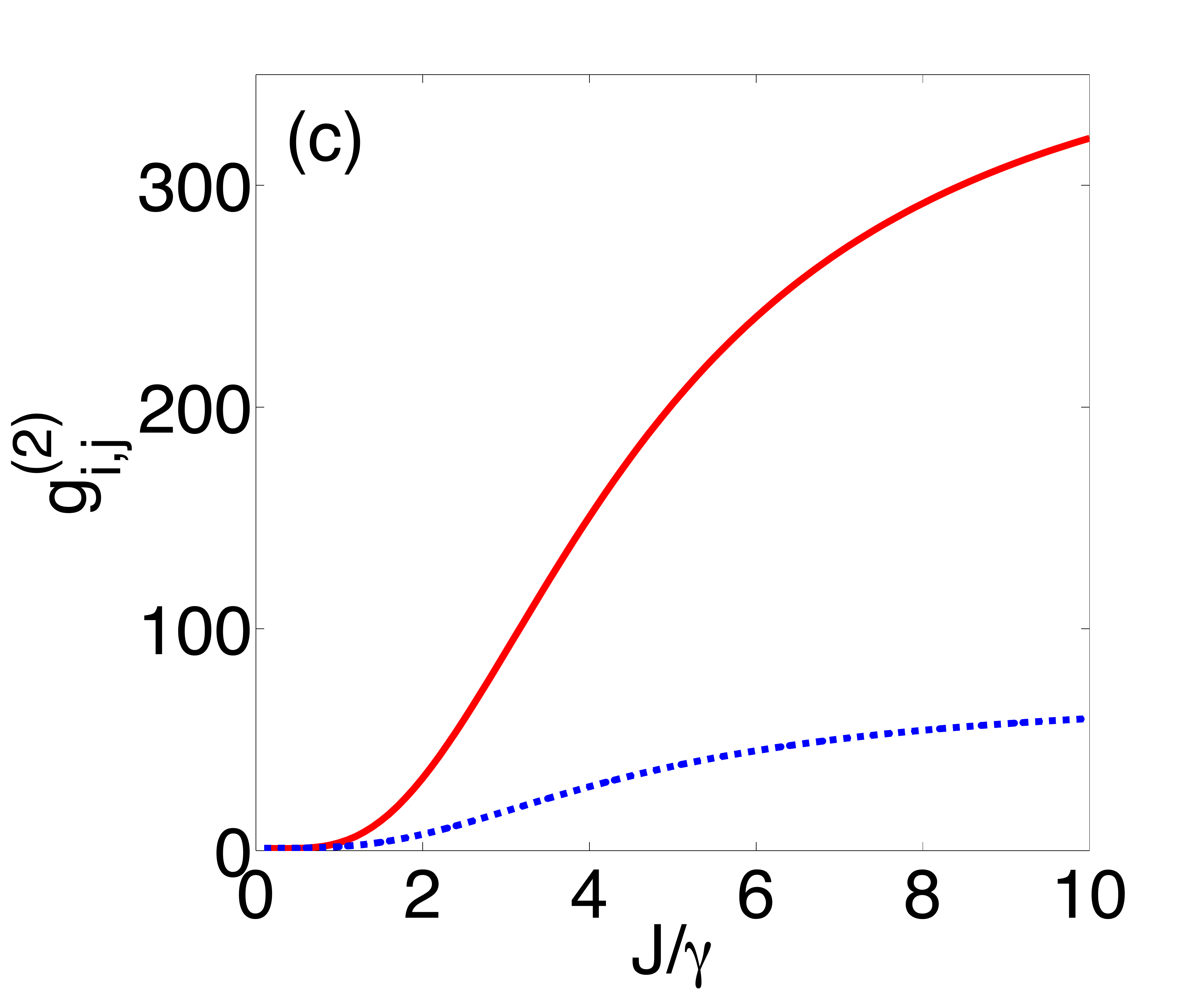} }
 \resizebox{0.42\columnwidth}{!}{%
 \includegraphics{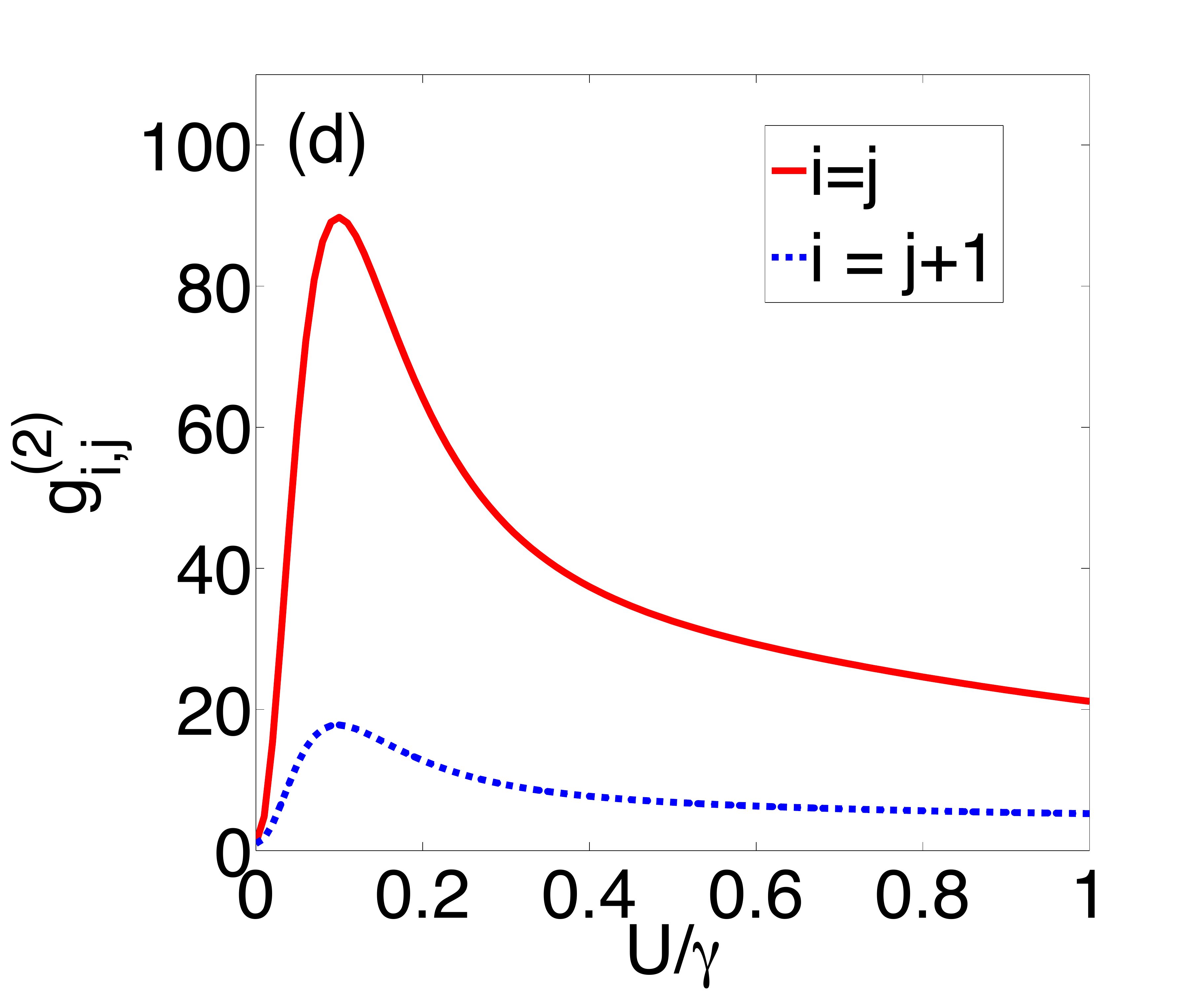} }
 \end{center}
\caption{On-site (solid red line) and non-local (blue dashed line) second order correlation function, as a function of the system parameters (in units of the loss rate $\gamma$), namely (a) the detuning $\Delta$, (b) the driving amplitude $F$, (c) the hopping term $J$, and (d) the on-site photon-photon interaction $U$. While not varied, the fixed parameters are $\Delta = 0$, $U = 0.1 \gamma$, $J = 3 \gamma$ and $F = 0.5 \gamma$.}
\label{fig:g2}       
\end{figure}

\begin{figure}
\begin{center}
\resizebox{0.25\columnwidth}{!}{
\includegraphics{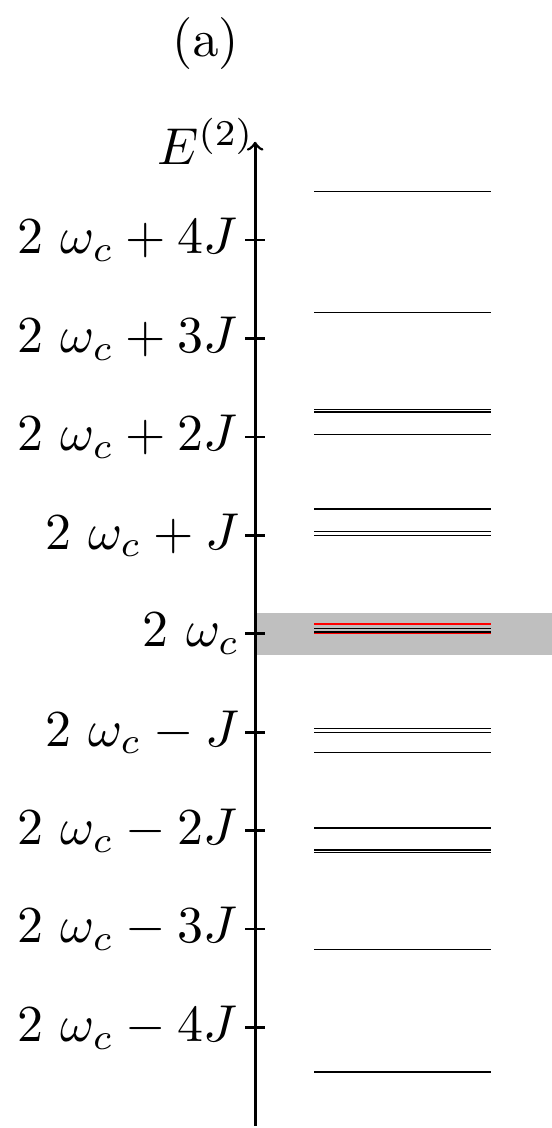} }
\resizebox{0.7\columnwidth}{!}{
\hspace{0.05\columnwidth}
\includegraphics{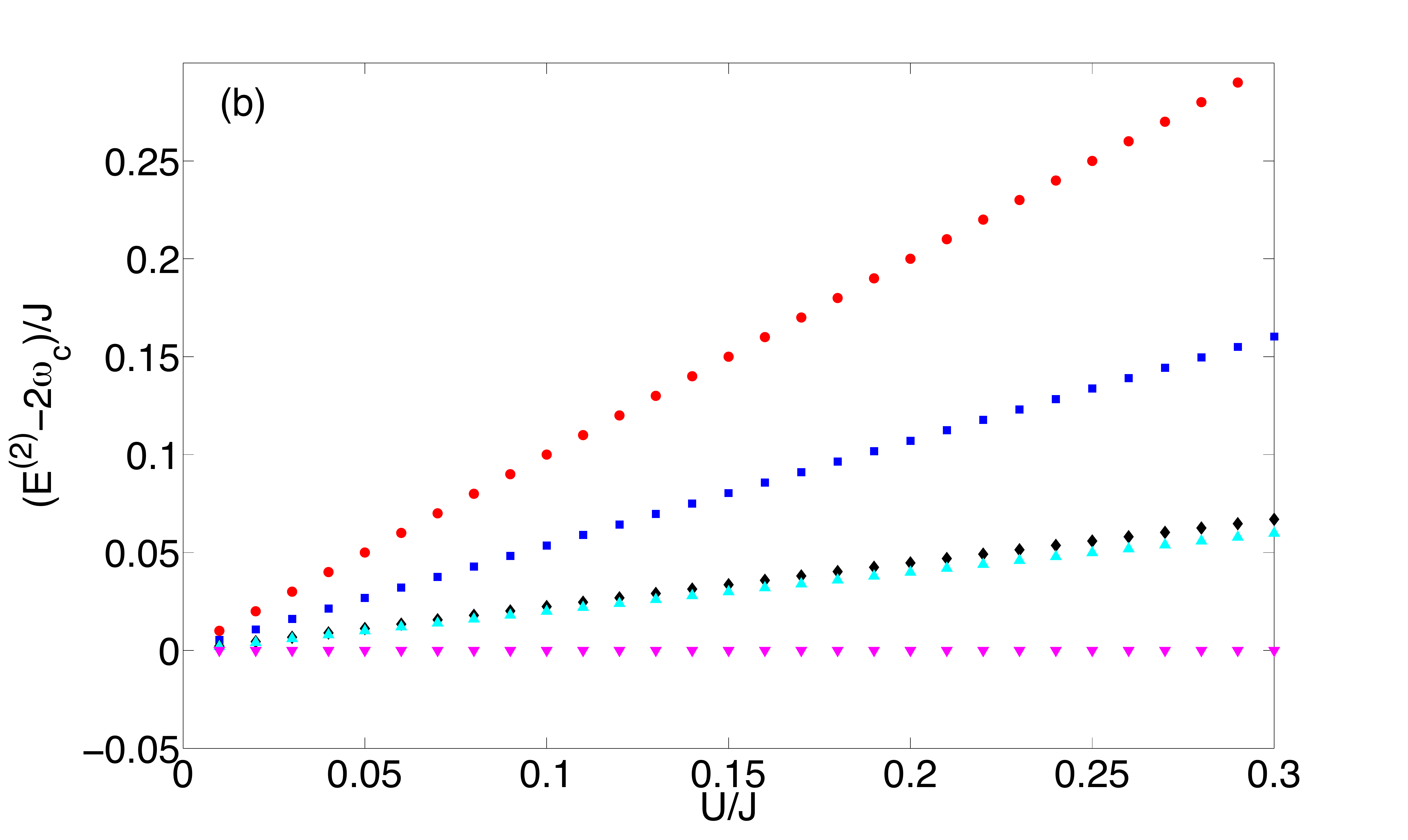} }
\end{center}
\caption{Energy spectrum of the two-photon eigenstates of the closed system. Panel (a): energy level structure of the two-photon states for $U = 0.1 \gamma$. The gray stripe highlights the energy range close to $E^{(2)} = 2 \omega_c$ (i.e. $2 (\omega_c - U) < E^{(2)} <  2 (\omega_c + U)$). The red lines indicates the levels corresponding to the two states in Eq. (\ref{eq:2PhotEigenstateA}-\ref{eq:2PhotEigenstateB}). Panel (b): energy of the five two-particle states on resonance for zero detuning in the limit of small non-linearity (each symbol corresponds to a different state), as a function of the nonlinearity $U$ (in units of the hopping strength $J$). }
\label{fig:spectrum2photons}       
\end{figure}

The hypothesis of the presence of strongly correlated photons at the dark sites is supported by the behavior of the second order correlation functions on the dark sites: $g^{(2)}_{i,j} = \average{\hat{b}_i^\dagger \hat{b}_j^\dagger \hat{b}_j \hat{b}_i}/n_b^2$. The results for the on-site ($i=j$) and the non-local ($i\ne j$) correlation functions $g^{(2)}_{i,j}$ are shown in Fig. \ref{fig:g2}, as a function of the different system parameters. They unambiguously show a deviation from the Gross-Pitaevskii prediction $g^{(2)}_{1,1} = g^{(2)}_{1,2} = 1$. The qualitative behavior of these correlation functions is similar to that obtained for the small system of three coupled cavities and for the extended Lieb lattice \cite{Casteels16Lieb}. Fig. \ref{fig:g2}-(a) shows the $g^{(2)}$ as a function of the detuning: it reaches its maximum value for $\Delta \simeq 0$, where the population of dark sites due to the uncorrelated coherent states is at minimum, while the correlated states play an important role. Interestingly, we notice that both the on-site and the non-local correlation functions are not symmetric around $\Delta = 0$ and their maximum is reached for a non-zero value of the detuning. These effects are due to the presence of the non-linearity which shifts the energy levels corresponding to the correlated states responsible for the large values of $g^{(2)}$ (see also Fig. \ref{fig:spectrum2photons}-(a) ). Fig. \ref{fig:g2}-(b) shows that correlations decreases as the pump intensity $F$ increases: indeed, as the driving term becomes dominant in the Hamiltonian, the system tends to a coherent state. Fig. \ref{fig:g2}-(c) reveals that the correlated states become more important as the hopping parameter $J$ is increased. Indeed, as $J$ gets larger, the energy separation between the flat and the dispersive bands increases, so that the contribution of the resonances at $\Delta = \pm \sqrt{5} J$  to the density of the dark sites gets smaller and smaller. Finally, fig. \ref{fig:g2}-(d) shows the dependence of $g^{(2)}$ on the nonlinearity $U$. In the regime of a linear system ($U = 0$), the laser drive pumps into the lattice coherent photons, for which $g^{(2)}_{i,i} = g^{(2)}_{i,j} = 1$. In the opposite limit of a large non-linearity, anti-bunching effects due to the increasing repulsion among the photons tend to decrease their mutual correlations (notice that in the hard-core limit $U \to \infty$, $g^{(2)}_{i,i} \to 0$). The strongly correlated states studied in this work, therefore, play a relevant role in the intermediate regime of small non-linearities, as indicated by the maximum value of the correlation functions which appears around $U/\gamma \simeq 0.2$.

In order to give a better insight about these correlated states, we study the spectrum of the two-photon eigenstates of the closed system, without driving and dissipation. From the diagonalization of the Hamiltonian in Eq. (\ref{eq:Hamiltonian}) with $F = 0$, we find 21 two-photon resonances of which the eigenenergies $E^{(2)}$ are presented in Fig \ref{fig:spectrum2photons}-(a) for a fixed value of the non-linearity ($U = 0.1 \gamma$). When a laser pump with frequency $\omega_p = \omega_c$ (i.e. in the regime of  detuning $\Delta = 0$) is applied to the lattice, this will tend to populate those states with energy close to the resonance $E^{(2)} = 2\omega_c = 2\omega_p$. We notice that, in the limit of small non-linearities, there are 5 two-photon states which satisfy the condition $\lim_{U \to 0} E^{(2)} = 2 \omega_c$ (see Fig. \ref{fig:spectrum2photons}-(b)). In particular, we focus our attention on two of them, i.e. the eigenstates corresponding to the energies $E = 2 \omega_c$ and $E = 2 \omega_c + U$. Written in the Fock basis representation $|n_{a_1},n_{b_1},n_{c_1},n_{a_2},n_{b_2},n_{c_2} \rangle$, these states are respectively:
\begin{eqnarray}
|\Psi_1 \rangle = & \frac{1}{\sqrt{6}} & \left( |2,0,0,0,0,0 \rangle - |0,2,0,0,0,0 \rangle + |0,0,2,0,0,0 \rangle \nonumber \right. \\
\label{eq:2PhotEigenstateA}
& & + \left. |0,0,0,2,0,0 \rangle - |0,0,0,0,2,0 \rangle + |0,0,0,0,0,2 \rangle \right) \\
|\Psi_2 \rangle = & \frac{1}{\sqrt{3}} & \left( |1,0,0,1,0,0 \rangle - |0,1,0,0,1,0 \rangle + |0,0,1,0,0,1 \rangle \right)
\label{eq:2PhotEigenstateB}
\end{eqnarray}
It is clear that $|\Psi_1 \rangle$ contains a non-zero amplitude for a double occupancy of the dark-sites, while $|\Psi_2 \rangle$ contains a non-zero amplitude for having one photon on each of the two $b$-sites. This shows that an occupation of these states is expected to have a large influence on the second order correlation functions on the dark sites, as indeed observed in Fig. \ref{fig:g2}. Therefore, we expect that these states contribute to the anomalous peak around $\Delta = 0$ in the dark site density.

\begin{figure}
\begin{center}
\resizebox{0.48\columnwidth}{!}{
\includegraphics{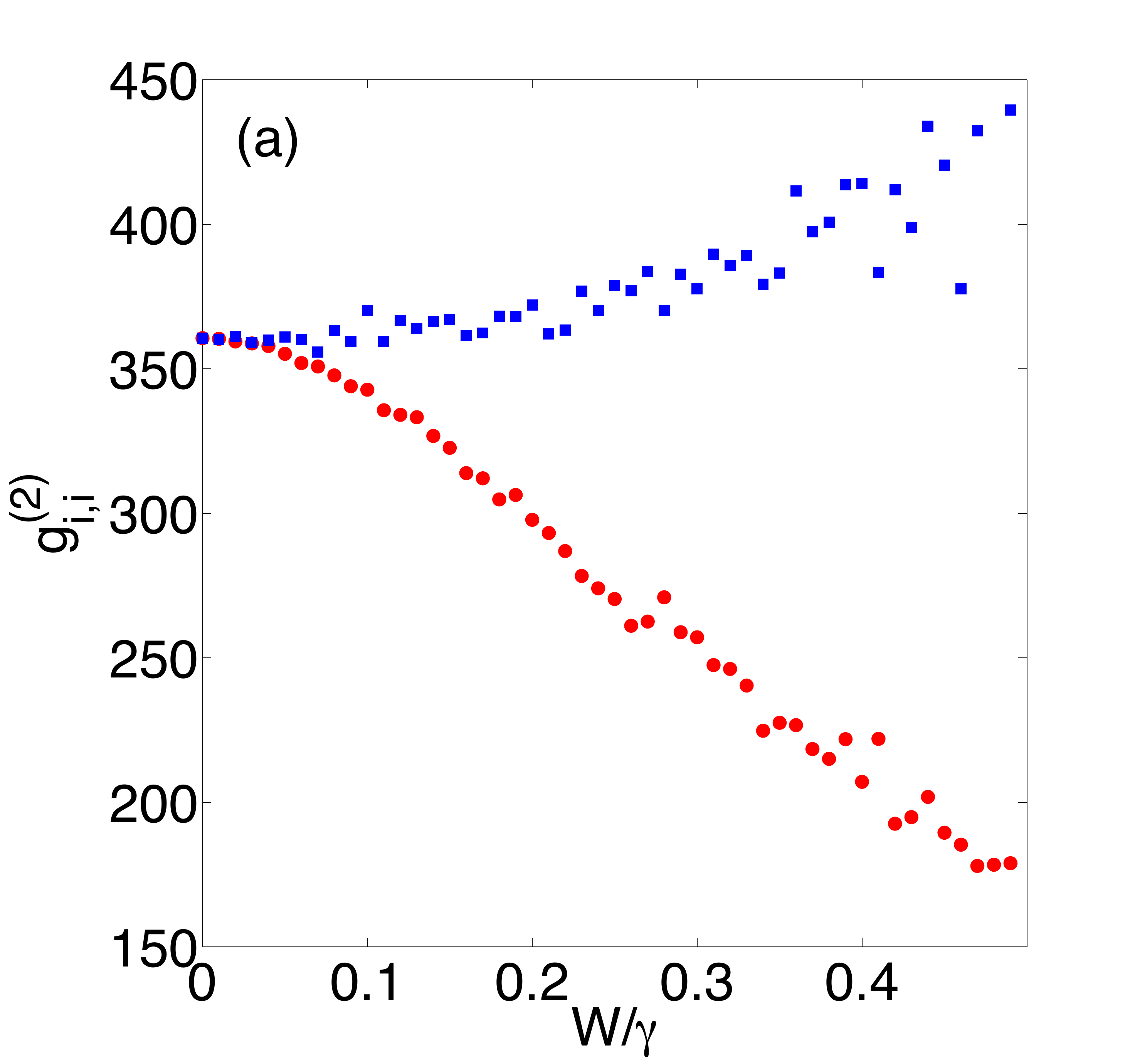}
}
\resizebox{0.48\columnwidth}{!}{
\includegraphics{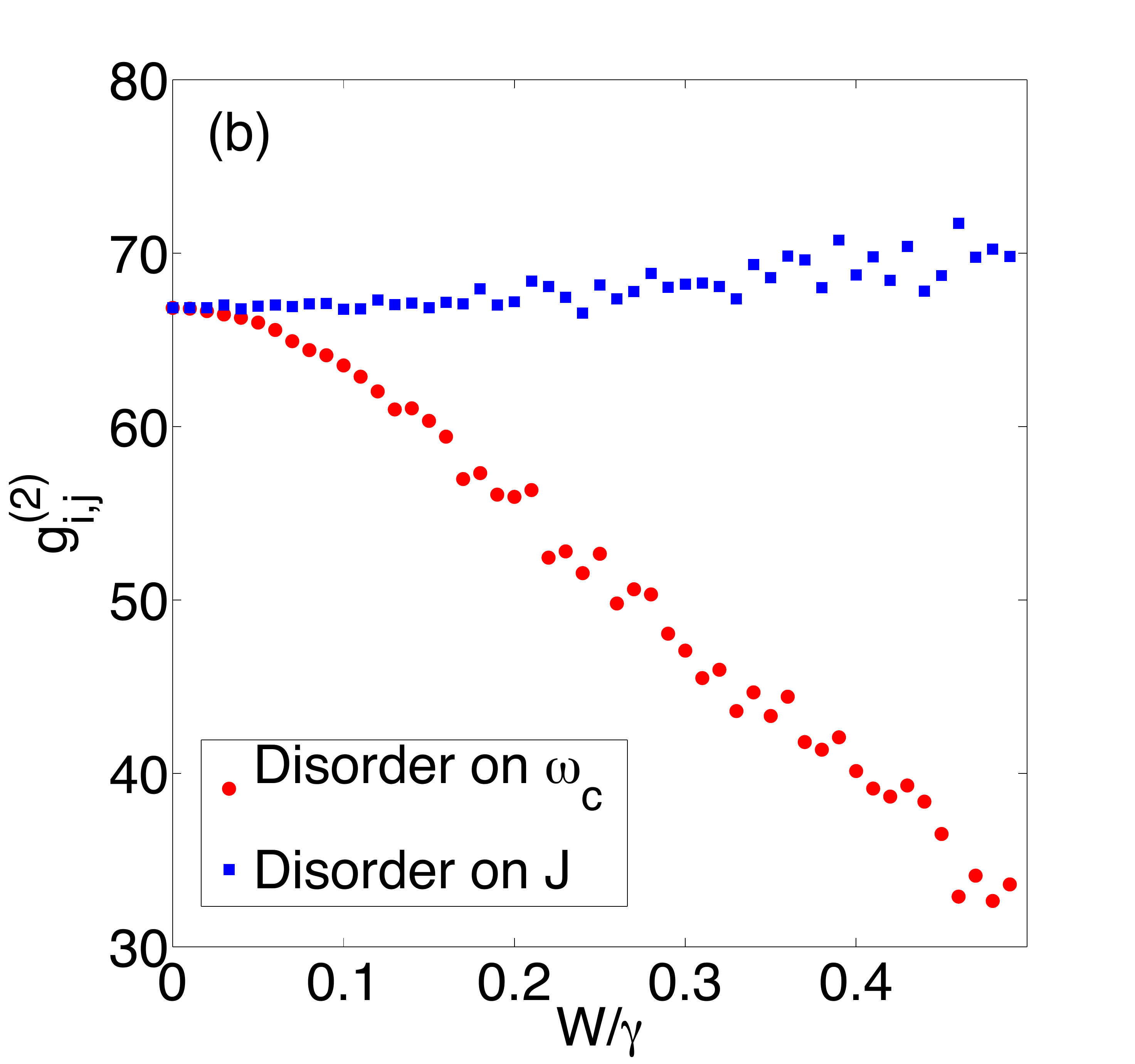}
 }
\end{center}\caption{Dependence of the on-site (a) and non-local (b) second order correlation function as a function of the disorder strength (in units of the loss rate $\gamma$). The different symbols represent the results with disorder on the cavity frequency $\omega_c$ (red circles) and with disorder on the hopping term $J$ (blue squares).}
\label{fig:disorder}       
\end{figure}

An important aspect that need to be understood is the effect of disorder on the multi-photon states, which is particularly relevant since quantum systems displaying flat energy bands are expected to be particularly sensitive to small perturbations. To perform this analysis, we solve the Master equation (Eq. \ref{eq:Master}) using a modified Hamiltonian, obtained by assuming that every cavity $s$ has a different frequency $\omega_{c,s} = \omega_c + W \xi_{s}$, where $\xi_{s}$ are random numbers homogeneously distributed in the interval $\left[ -\frac{1}{2} , \frac{1}{2}\right]$ and $W$ is a parameter denoting the strength of the disorder. In fig. \ref{fig:disorder}, we show the dependence of both the on-site (Fig. \ref{fig:disorder}-(a)) and the non-local correlation function (Fig. \ref{fig:disorder}-(b)) as a function of the disorder strength $W$: the results are obtained by averaging over 200 different realizations of the disorder, obtained by changing the set of the random numbers $\xi_{s}$. The decreasing of $g^{(2)}_{i,j}$ as a function of $W$,  both for $i=j$ and $i\ne j$, indicates that local disorder has a detrimental effect on the strong correlations displayed at dark sites. We have also analyzed the effect of hopping disorder by considering different coupling between the cavities. More precisely, we modify the Hamiltonian of the lattice by writing the hopping strength between cavities $s$ and $t$ as $J_{s,t} = J + W \xi_{s,t}$ and we calculate the second-order correlation functions as a function of the disorder parameter $W$ (as in the previous case, the quantities $\xi_{s,t}$ are random numbers homogeneously distributed in the interval $\left[ -\frac{1}{2} , \frac{1}{2}\right]$). The results obtained are also displayed in fig. \ref{fig:disorder}: they show that, contrarily to the disorder on the cavity frequencies, the hopping disorder hardly affects the strong correlations among photons.

%

%
%
\section{Conclusions}\label{sec:conclusions}
We have studied the driven-dissipative physics of a frustrated lattice composed of six coupled photonic cavities, which has the same geometry as the 1D Lieb lattice. The single-particle energy spectrum of this lattice displays a flat band associated to localized states having dark sites with zero photon occupation.

Using an innovative truncation method of the infinite Hilbert space of the quantum system, which is based on the global correlation order rather than on the maximum number of photon per sites, we have been able to efficiently calculate the non-equilibrium steady state and to estimate the expectation values of the relevant observables.

We have shown that, when the driving is at resonance with the flat energy band, the occupation of multi-photon states strongly affects the properties of the dark sites. These latter, indeed, show an additional photonic population with strong correlations, even in regimes of weak photon-photon interaction. These correlation effects are strongly reduced in presence of a disorder on the cavity frequencies, but seem insensitive to a disorder on the coupling between different sites.

\section*{Acknowledgements}
We acknowledge financial support from ERC, through the Consolidator Grant "CORPHO" (No. 616233).

\end{document}